\documentclass[11pt,onecolumn]{IEEEtran}
\usepackage[dvips]{color}
\usepackage{epsf}
\usepackage{times}
\usepackage{epsfig}
\usepackage{graphicx}
\usepackage{url}
\usepackage{amsmath}
\usepackage{amssymb}
\usepackage{amsxtra}
\usepackage{algorithmic}
\usepackage{algorithm}
\usepackage{multirow}
\usepackage{cite}
\usepackage{enumerate}
\usepackage{wrapfig}

\newtheorem{definition}{\bf Definition}

\newlength{\aligntop}
\newlength{\alignbot}
\makeatletter

\IEEEoverridecommandlockouts
\begin{document}

\title{Game Theoretic Methods for the Smart Grid}

\author{\authorblockN{ Walid Saad$^\textbf{1}$, Zhu Han$^\textbf{2}$, H. Vincent Poor$^\textbf{3}$, and Tamer Ba\c{s}ar$^\textbf{4}$\\} \authorblockA{\small
$^\textbf{1}$ Electrical and Computer Engineering Department, University of Miami, Coral Gables, FL, USA, email: \url{walid@miami.edu}\\
$^\textbf{2}$ Electrical and Computer Engineering Department, University of Houston, Houston, TX, USA, email: \url{zhan2@mail.uh.edu}\\
$^\textbf{3}$ Electrical Engineering Department, Princeton University, Princeton, NJ, USA, email: \url{poor@princeton.edu}\\
$^\textbf{4}$ Coordinated Science Laboratory, University of Illinois at Urbana-Champaign, USA, email: \url{basar1@illinois.edu}}\vspace{-3.8cm}\thanks{This work was supported in part by the U.S. National Science Foundation
under Grants CNS-09-05398, CNS-09-05086, CCF-10-16671, DMS-11-18605, CNS-1117560, ECCS-1028782, CNS-0953377,  and CNS-0905556, in part by the Qatar National Research Fund under grant NPRP 4 - 347 - 2 - 127, and in part by  the U.S. Department of Energy (DOE) under grant number DE-SC0003879 and by
the U.S. Air Force Office of Scientific Research (AFOSR) under grant number
MURI FA9550-09-1-0249.}}
\date{}
\maketitle

\thispagestyle{empty}
\vspace{5cm}
\begin{abstract}
\setlength{\baselineskip}{20pt}
The future smart grid is envisioned as a large-scale cyber-physical system encompassing advanced power, communications, control, and computing technologies. In order to accommodate these technologies, it will have to build on solid mathematical tools that can ensure an efficient and robust operation of such heterogeneous and large-scale cyber-physical systems. In this context, this paper is an overview on the potential of applying game theory for addressing relevant and timely open problems in three emerging areas that pertain to the smart grid: micro-grid systems, demand-side management, and communications. In each area, the state-of-the-art contributions are gathered and a systematic treatment, using game theory, of some of the most relevant problems for future power systems is provided. Future opportunities for adopting game theoretic methodologies in the transition from legacy systems toward smart and intelligent grids are also discussed. In a nutshell, this article provides a comprehensive account of the application of game theory in smart grid systems tailored to the interdisciplinary characteristics of these systems that integrate components from power systems, networking, communications, and control.
\end{abstract}

\vspace{-0.3cm}
\indent \indent 
\newpage
 \setcounter{page}{1} \setlength{\baselineskip}{18pt}
\section{Introduction and motivation}\vspace{-0.2cm}
 The smart grid is envisioned to be a large-scale cyber-physical system that can improve the efficiency, reliability, and robustness of power and energy grids by integrating advanced techniques from various disciplines such as power systems, control, communications, signal processing, and networking.  Inherently, the smart grid is a power network composed of intelligent nodes that can operate, communicate, and interact, autonomously, in order to efficiently deliver power and electricity to their consumers. This heterogeneous nature of the smart grid motivates the adoption of advanced techniques for overcoming the various technical challenges at different levels such as design, control, and implementation.

In this respect, \emph{game theory} is expected to constitute a key analytical tool in the design of the future smart grid, as well as large-scale cyber-physical systems. Game theory is a formal analytical as well as conceptual framework with a set of mathematical tools enabling the study of complex interactions among independent rational players. For several decades, game theory has been adopted in a wide number of disciplines ranging from economics and politics to psychology \cite{GT01}. More recently, game theory has also become a central tool in the design and analysis of communication systems~\cite{MC01}.

The proliferation of  advanced technologies and services in smart grid systems implies that disciplines such as game theory will naturally become a prominent tool in the design and analysis of smart grids. In particular, there is a need to deploy novel models and algorithms that can capture the following characteristics of the emerging smart grid: (i)- the need for distributed operation of the smart grid nodes for communication and control purposes, (ii)- the heterogeneous nature of the smart grid which is typically composed of a variety of nodes such as micro-grids, smart meters, appliances, and others, each of which having different capabilities and objectives, (iii)- the need for efficiently integrating advanced techniques from power systems, communications, and signal processing, and (iv)- the need for low-complexity distributed algorithms that can efficiently represent competitive or collaborative scenarios between the various entities of the smart grid. In this context, game theory could constitute a robust framework that can address many of these challenges.

In this paper, we aim to provide a systematic treatment of applying game theory in smart grids. In particular, our objectives are three-fold: (i)- to provide a comprehensive description of existing game theoretic applications in smart grid networks, (ii)- to identify key open problems in smart grid networks that are bound to be addressed using game theory, and (iii)- to pinpoint the main game theoretic tools that can be adopted for designing the smart grid. Thus, we seek to provide the reader with a clear picture of the underlying strengths and challenges of adopting classical and novel game theoretic techniques within the context of the smart grid.

The remainder of this paper is organized as follows. In Section~\ref{sec:review}, we briefly review some key game theoretic concepts. Then, in Section~\ref{sec:mg}, we study the applications of game theory in micro-grids while in Section~\ref{sec:dsm}, we discuss how game theory can be applied to demand-side management. In Section~\ref{sec:com}, we survey the applications of game theory for smart grid communications. Finally, a summary is provided in Section~\ref{sec:conc}.\vspace{-0.3cm}


\section{Review of Fundamental Game Theoretic Concepts}\label{sec:review}
\subsection{Introduction and Basic Game-Theoretic Concepts}
Game theory is a mathematical framework that can be divided into two main branches: noncooperative game theory and cooperative game theory. \emph{Noncooperative game theory} can be used to analyze the strategic decision making processes of a number of independent entities, i.e., players, that have partially or totally conflicting interests over the outcome of a decision process which is affected by their actions. Essentially, noncooperative games can be seen as capturing a distributed decision making process that allows the players to optimize, without any coordination or communication, objective functions coupled in the actions of the involved players. We note that the term noncooperative  does not always imply that the players do not cooperate, but it means that, any cooperation that arises must be self-enforcing with no communication or coordination of strategic choices among the players. To this end, one of the recently emerging areas is that of designing incentives to enforce cooperation in a noncooperative setting, such as in \cite{NEW00} and \cite{NEW01}.

\subsubsection{Basics of Noncooperative Game Theory} Noncooperative games can be grouped into two categories: static games and dynamic games. Static games are games in which the notions of time or information do not affect the action choices of the players. Thus, in a static setting, a noncooperative game can be seen as a one-shot process in which the players take their actions only once (simultaneously or at different points in time). In contrast, dynamic games are games in which the players have some information
about each others' choices, can act more than once, and time has a central role in the decision making. For static games, one general definition is the following\footnote{This is known as the \emph{strategic or normal form} of a game~(see \cite{GT01} for more details).}:
\begin{definition}
A static noncooperative game is defined as a situation that involves three components: the set of players $\mathcal{N}$, the action sets $(\mathcal{A}_i)_{i\in \mathcal{N}}$, and the utility functions $(u_i)_{i\in \mathcal{N}}$. In such a noncooperative game, each player $i$ wants to choose an action $a_i \in \mathcal{A}_i$ so as to optimize its utility function $u_i(a_i,\boldsymbol{a}_{-i})$ which depends not only on player $i$'s action choice $a_i$ but also on the vector of actions taken by the other players in $\mathcal{N}\setminus \{i\}$, denoted by $\boldsymbol{a}_{-i}$.
\end{definition}

Note that, when the game is dynamic, one needs to also define, as part of the game, additional components such as information sets, time, or histories (i.e., sets of past actions) which are usually reflected in the utility functions. We note that the notion of action coincides with that of a \emph{strategy} in static games while in dynamic games strategies are defined, loosely, as functions of the information available to each player (the interested reader is referred to \cite{GT01} for more details). For the scope of this paper, we will use interchangeably the terms action and strategy, unless an explicit distinction is required. The strategy choices of the players can be made either in a deterministic manner, i.e., \emph{pure strategies}, or by following a certain probability distribution over the action sets $(\mathcal{A}_i)_{i\in \mathcal{N}}$, i.e., \emph{mixed strategies}.

\subsubsection{Solution Concept} The objective of noncooperative game theory is to provide algorithms and techniques suitable for solving such optimization problems and characterizing their outcome, notably when the players are making their action choices noncooperatively (and independently), i.e., without any coordination or communication. One of the most important solution concepts for game theory in general and noncooperative games in particular is that of a \emph{Nash equilibrium}. The Nash equilibrium characterizes a state in which no player $i$ can improve its utility by changing \emph{unilaterally} its strategy, given that the strategies of the other players are fixed. For a static game, the Nash equilibrium in pure strategies can be formally defined as follows:
\begin{definition}\label{def:nash}
A \emph{pure-strategy Nash equilibrium} of a static noncooperative game is a vector of actions $\boldsymbol{a}^* \in \mathcal{A}$ ($\mathcal{A}$ is the Cartesian product of the action sets) such that $\forall i \in \mathcal{N}$, the following holds:
\begin{equation}
u_i(a_i^*,\boldsymbol{a}_{-i}^*) \ge u_i(a_i,\boldsymbol{a}_{-i}^*),\ \forall a_i \in \mathcal{A}_i.
\end{equation}
\end{definition}

The Nash equilibrium serves as a building block for many types of noncooperative games. This solution concept has both advantages and drawbacks. One of its main advantages is that it characterizes a stable state of a noncooperative game in which no player $i \in \mathcal{N}$ can improve its utility by \emph{unilaterally} changing its action $a_i$ given that the actions of the others are fixed at $\boldsymbol{a}_{-i}^*$. This state can often be reached by the players in a distributed manner and with little coordination~\cite{GT01}. However, the Nash equilibrium also has some drawbacks. For instance, even in finite games (that is games where each player has a finite action set), a Nash equilibrium is only guaranteed to exist in \emph{mixed strategies}\footnote{In mixed strategies, a Nash equilibrium is defined similar to Definition~\ref{def:nash} with the strategies being a vector of probability distributions over the action sets.}. Also, a noncooperative game can have multiple Nash equilibria and, thus, selecting an efficient and desirable Nash equilibrium is a challenging topic, notably when applied to practical systems such as networks~\cite{MC01}. Nonetheless, several metrics such as the price of anarchy or the price of stability can be used to study the efficiency the Nash equilibrium such as in~\cite{NEW02}. Moreover, the Nash equilibrium concept can be complemented and extended using many other game theoretic techniques such as pricing so as to provide suitable solutions for noncooperative games~\cite{GT01}.

\subsubsection{Cooperative Games} In noncooperative games, it is assumed that the players are unable to coordinate or communicate with one another directly. However, for games in which the players are allowed to communicate and to receive side payments (e.g., share utilities), it may be of interest to adopt fully cooperative approaches. In this respect, \emph{cooperative game theory} provides frameworks that can answer one pertinent question: ``What happens when the players can communicate with one another and decide to cooperate?''. Cooperative games allow to investigate how one can provide an incentive for independent decision makers to act together as one entity so as to improve their position in the game. For example, in politics, different parties may decide to merge or coalesce into a cooperative group so as to improve their chances in obtaining a share of the power. Cooperative game theory encompasses two parts: Nash bargaining and coalitional game. Nash bargaining deals with situations in which a number of players need to agree on the terms under which they cooperate while coalitional game theory deals with the formation of cooperative groups or coalitions. In essence, cooperative game theory in both of its branches provides tools that allow the players to decide on whom to cooperate with and under which terms given several cooperation incentives  and fairness rules. A detailed treatment of cooperative game theory can be found in \cite{MC01}.

\subsection{Learning in Games}
While studying the efficiency of an equilibrium is central to game-theoretic design, another important aspect is to develop learning algorithms that enable the players to reach a certain desired game outcome. In fact, choosing the desired equilibrium is a challenging topic that has warranted many recent research efforts~\cite{LEARN00,FL98,GT01,NEW02,LEARN04,LEARN07}. To reach a certain equilibrium, the players must follow well-defined rules that enables them to observe the current game state and make a decision on their strategy choices. Essentially, a learning scheme is an iterative process in which each iteration involves three key steps performed by every player~\cite{LEARN00}: (i)- observing the environment and current game state, (ii)- estimating the prospective utility, and (iii)- updating the strategy based on the observations.

Numerous learning algorithms have been proposed in the literature~\cite{LEARN00,FL98,GT01,NEW02,LEARN04,LEARN07}. The simplest of such algorithms is the so-called \emph{best response dynamics} which is an iterative process in which, at each iteration, a player selects the strategy that maximizes its utility, i.e., its best response strategy. Several variants of this process exist. One of the advantages of a best response algorithm is its simple implementation, however, it suffers from several drawbacks. First, a best response process is only guaranteed to converge to an equilibrium for certain types of utility functions~\cite{FL98}. Second, best response dynamics are highly sensitive to the initial conditions and any changes in these conditions could lead to different equilibria. Third, adopting a best response approach does not always guarantee convergence to an efficient equilibrium~\cite{FL98}.

In this respect, several more advanced algorithms have been studied for learning the equilibria of a game theoretic model. While a detailed treatment of such algorithms is outside the scope of this survey\footnote{For more details, the interested reader is referred to the abundant literature on learning\cite{LEARN00,FL98,GT01,NEW02,LEARN04,LEARN07}}, we provide the following summary that can guide the interested reader towards the relevant literature:
\begin{itemize}
\item \emph{Fictitious play:} Fictitious play refers to a family of iterative learning algorithms in which, at each iteration, each player is able to observe the actions of all other players and compute the empirical frequency with which it chooses a certain action. Having estimated the empirical frequency, each player can subsequently select its optimal strategy, in a given iteration. It is shown that for some special types of games such as zero-sum games, fictitious play always converges to a Nash equilibrium~\cite{LEARN00}. Several recent contributions have also proposed many enhancements to fictitious play algorithms~\cite{LEARN00,FL98,LEARN04}.
\item \emph{Regret matching:} Fictitious play and best response algorithms are based on the idea that, at each iteration, a player attempts to maximize its utility. In contrast, regret matching is a type of learning algorithms in which the players attempt to minimize their regret from using a certain action, i.e., the difference between the utility of always playing a certain action and the utility that they achieved by playing their current strategy. An in-depth treatment of regret matching algorithms is found in \cite{LEARN00,FL98,LEARN07}.
\item \emph{Other learning schemes:} Many other types of learning schemes such as reinforcement learning or stochastic learning are also used in various game-theoretic scenarios in order to find a desirable state of the system~\cite{LEARN00}.
\end{itemize}

Clearly, learning is an integral part of game theory and it lies at the heart of designing stable and efficient models.

\subsection{Game Theory in the Smart Grid: Potential and Challenges}
Within the context of smart grids, the applications of noncooperative games and of learning algorithms are numerous. On the one hand, noncooperative games can be used to perform distributed demand-side management and real-time monitoring or to deploy and control micro-grids. On the other hand, economical factors such as markets and dynamic pricing are an essential part of the smart grid. In this respect, noncooperative games provide several frameworks ranging from classical noncooperative Nash games to advanced dynamic games which enable to optimize and devise pricing strategies that adapt to the nature of the grid. Several practical noncooperative game examples in the smart grid are treated, in details, in the remainder of this paper.

In smart grids, with the deployment of advanced networking technologies, it is often possible to enable a limited form of communication between the nodes which paves the way for introducing cooperative game-theoretic approaches. In fact, the integration of power, communication, and networking technologies in future grids opens up the door for several prospective applications in which smart grid nodes can cooperate so as to improve the robustness and efficiency of the grid. One simple example would be to apply cooperative game theory in order to study how relaying can be performed in a large-scale smart grid network so as to improve the efficiency of the communication links between smart grid elements. Other cooperative game applications are also possible as seen later in this survey.

Clearly, game-theoretic approaches present a promising tool for the analysis of smart grid systems. Nonetheless, the advantages of applying distributed game-theoretic techniques in any complex system such as the smart grid are accompanied by key technical challenges. First, one of the underlying assumptions in classical game-theoretic designs is that the players are \emph{rational}, i.e., each player makes its strategy choice so as to optimize its individual utility and, thus, conform with some notion of equilibrium play. In practical control systems such as the smart grid, as the individual nodes of the system interact and learn their strategies, one ore more nodes might deviate from the intended play and make non-rational decisions, i.e., choose unintended strategies, due to various factors such a failure or delay in learning. These inaccurate strategy choices can eventually lead to a non-convergence to the desired equilibrium and, hence, impact the overall control system stability. The impact of such bad decisions becomes more severe in practical deployments in which the smallest perturbation to the system stability can lead to outages or other detrimental consequences. As a result, when designing game-theoretic models for the smart grid, it is imperative to emphasize robustness in the model and algorithm design.

Fortunately, the analytical framework of game theory presents several remedies for this difficulty~\cite{LEARN00,FL98,GT01,NEW02,LEARN04,LEARN07}. First, classical game-theoretic techniques already provide a myriad of analytical approaches for ensuring robustness against errors. These approaches include the concept of a perturbed equilibrium (e.g., trembling hand equilibrium or strong time consistency)~\cite{GT01,FL98} and games with imperfect information or imperfect observability~\cite{GT01}, among others. Second, one common approach for overcoming the instability resulting from bad decisions is to design learning algorithms in which the players are allowed to ``experiment'', i.e., choose unintended strategies. The essence of these algorithms is to turn the bad decisions into an opportunity to improve the efficiency of the reached equilibrium. These learning schemes fall under the umbrella of learning by experimentation which has received significant attention recently in game theory and multi-agent learning (see \cite{LEARN00} and references therein). These approaches have also been recently shown to perform well in practical communication systems~\cite{LEARN07,MC01} and, hence, it is natural to leverage their use into practical control systems such as the smart grid, so as to avoid erroneous decisions and instability. Third, the emerging field of game theory with bounded rationality provides a set of tools and concepts (such as the limited foresight equilibrium~\cite{TRIAL02} or the logit equilibrium~\cite{LOGIT}) for designing distributed optimization techniques that are robust to possible non-rational decisions or deviations from the players~\cite{BR00}. These techniques can certainly be leveraged so as to address the issue of wrongful strategy choices during game-theoretic designs.

Beyond errors in the decision making process, game-theoretic designs in control systems also face other important challenges such as avoiding the possibility of cheating (in power market auctions for example) and adapting the learning  process to environmental variations, among others. To address these issues several new techniques such as strategy-proof auctions or advanced learning techniques have been proposed in \cite{LEARN00,FL98}.

In summary, game theory has a strong potential for addressing several emerging problems in smart grid systems, as detailed in the remainder of this article. However, in order to reap the benefits of game-theoretic designs, one must address some of the aforementioned challenges, notably when transferring these designs from a simulated environment to a practical system. Certainly, to do so, constant feedback between theory and practice is needed. The recent results and advances in game-theoretic designs in practical wireless and communication systems~\cite{MC01,LEARN07} corroborate the promising potential of deploying these designs in the future smart grid and could serve as a first step towards practical adoption of game theory in power systems.

In the remainder of this paper, we explore the applications of game theory in three key smart grid areas: (i)- micro-grid distribution networks, (ii)- demand-side management, and (iii)- communication protocols. We provide several carefully drawn examples on applying game theory in each one of these areas, and, then, we shed a light on future opportunities and key challenges. \vspace{-0.3cm}

\section{Game Theory in Micro-Grid Distribution Networks}\label{sec:mg}\vspace{-0.1cm}
In this section, we provide an overview on the deployment of micro-grids in future power systems and a survey on potential applications of game theory in micro-grids. Then, we study, in detail, the use of cooperative games for enabling energy trading between micro-grids and the use of noncooperative games for load and source control. We conclude with insights on future game-theoretic approaches for micro-grid distribution networks.\vspace{-0.1cm}
\subsection{Introduction to Micro-grids}
A power grid system can, in general, be divided into two main phases: electric power transmission and electric power distribution~\cite{MC00}. Electric power transmission or high-voltage transmission deals with the transmission of the energy generated at the power plants (i.e., the transfer of energy, over transmission lines, to substations that service some geographical areas). In contrast, electric power distribution is the last stage for delivering electricity in which the distribution network carries the electricity received at a substation and, subsequently, delivers it to the consumers' premises. A step-down transformer serves as a crossing point between the transmission and the distribution networks. This transformer serves to lower the high-voltage arriving from the transmission network so as to allow the distribution network to operate in low (or medium) voltage ranges.

In the past decade, researchers have been investigating the possibility of having groups of controllable loads and sources of energy at the distribution network side of a power grid~\cite{MG00}. In this context, the concept of a \emph{micro-grid} is defined as a networked group of distributed energy sources such as solar panels or wind turbines located at the distribution network side and which can provide energy to small geographical areas. The network of micro-grids is envisioned to operate both in conjunction with the grid as well as autonomously in isolated mode (known as the \emph{island} mode)~\cite{MG00}. In this respect, controlling the operation of the micro-grids and integrating them in the smart grid introduces several technical challenges that need to be addressed so as to ensure an efficient and reliable grid operation.

In classical power grids, it is common to optimize the system by defining and solving a system-level optimization problem based on a centralized objective function. However, in the presence of micro-grids, it is of interest to define a specific objective function for each micro-grid. This is mainly due to the heterogeneous nature of the micro-grid network which often consists of different components such as electric cars, energy storage devices (e.g. batteries), diesel generators, wind turbines, and solar farms. To this end, it is only natural that one adopts distributed analytical techniques such as game theory so as to control and optimize smart grid systems that encompass a micro-grid distribution network. This is further motivated by the vision of autonomous micro-grids which can act and react to various variables in the power grid system. In this respect, several open problems in micro-grids can be treated using game theory such as in \cite{MGA03,MGA00,WS03,MGA01,MG04,MG05,MG06,WS01,MGA02} (and references therein).

To give more insights on these open problems and on game-theoretic micro-grid designs, in what follows, first, we provide a step-by-step tutorial on how cooperative game theory can enable cooperative energy exchange in micro-grids. Then, we overview the use of noncooperative games for modeling the interactions between loads and sources in micro-grids. This section is concluded with a brief overview on other existing game-theoretic techniques in micro-grid design as well as with a discussion on the future outlook of game theoretic applications in micro-grids.\vspace{-0.3cm}

\subsection{A Game Theoretic Model for Cooperative Energy Exchange}\label{sec:coal}\vspace{-0.1cm}
\subsubsection{Cooperative Energy Exchange Model}
In existing power systems,  consumers are serviced by a main electricity grid that delivers the power over the transmission lines to a substation which, in turn, delivers the power over the low-voltage distribution network. In the presence of micro-grids, it is desirable to allow the micro-grids to service some small geographical areas or groups of consumers, so as to relieve the demand on the main grid. However, the intermittent generation of certain micro-grids coupled with the unpredictable nature of the consumers' demand implies that customers serviced solely by a micro-grid may, at certain points in time, become in need for extra energy from other sources. Typically, this extra energy need can be provided by the main power grid. 

The future smart grid is envisioned to encompass a large number of micro-grid elements. Hence, whenever some micro-grids have an excess of power while others have a need for power, it might be beneficial for these micro-grids (and their consumers) to exchange energy among each other instead of requesting it from the main grid. The advantage of such a local exchange is two-fold: (i)- energy exchange between nearby micro-grids can significantly reduce the amount of power that is wasted during the transmission over the distribution lines and (ii)- performing a local exchange of energy contributes further to the autonomy of the micro-grid system while reducing the demand and reliance on the main electric grid. Thus, it is of interest to devise a scheme that enables such a local energy trade between micro-grid elements that are in need of energy, i.e., \emph{buyers} and micro-grids that have an excess of energy to transfer, i.e., \emph{sellers}. To this end, as shown in \cite{WS01}, one can use cooperative games to introduce a cooperative energy exchange mechanism in future grids.

Consider a distribution network composed of a single substation which is connected to the main grid as well as to $N$ micro-grids in the set $\mathcal{N}$. Each micro-grid $i \in \mathcal{N}$ services a certain demand (e.g., a group of consumers or a small area) and the difference between its generation and demand is captured by a variable $Q_i$ which can be considered as random (due to the random nature of renewable energy generation and/or consumers' behavior). At a given period of time, depending on the consumers' demand and power generation, a certain  micro-grid $i \in \mathcal{N}$ (or its served area) may have either a surplus of power ($Q_i >0$) to sell or a need  to acquire power to meet its demand\footnote{Micro-grids who are able to exactly meet their demand, i.e., $Q_i=0$, do not participate in any energy exchange. However, they may remain connected to the larger grid in order to maintain frequency stability.}($Q_i < 0$).

 In the absence of storage and cooperation, each micro-grid $i \in \mathcal{N}$ exchanges (or acquires) the amount of power $Q_i$ with the main smart grid using the main substation. This transfer of power is accompanied by a power loss over the distribution lines inside the micro-grid network.  In this setting, by focusing on these power losses, the noncooperative utility of any micro-grid $i$ can be expressed as the total power loss over the distribution line due to the power transfer~\cite{WS01}:
\begin{equation}\label{eq:ncutil}
u(\{i\})= - w_i P_{i0}^{\textrm{loss}} ,
\end{equation}
where $P_{i0}^{\textrm{loss}}$ is power lost during power exchange between $i$ and the substation, $w_i$ is the price paid by $i$ per unit of power loss, and the minus sign is inserted to turn the problem into a maximization. The power loss $P_{i0}^{\textrm{loss}}$ is a function of several factors such as the distance between the micro-grid and the substation (due to the resistance), the power transfer voltage, the amount of power $Q_i$ that is being transferred, as well as the losses at the transformers of the substation.

\begin{figure}[t!]
  \begin{center}
   \vspace{-0.2cm}
    \includegraphics[width=10cm]{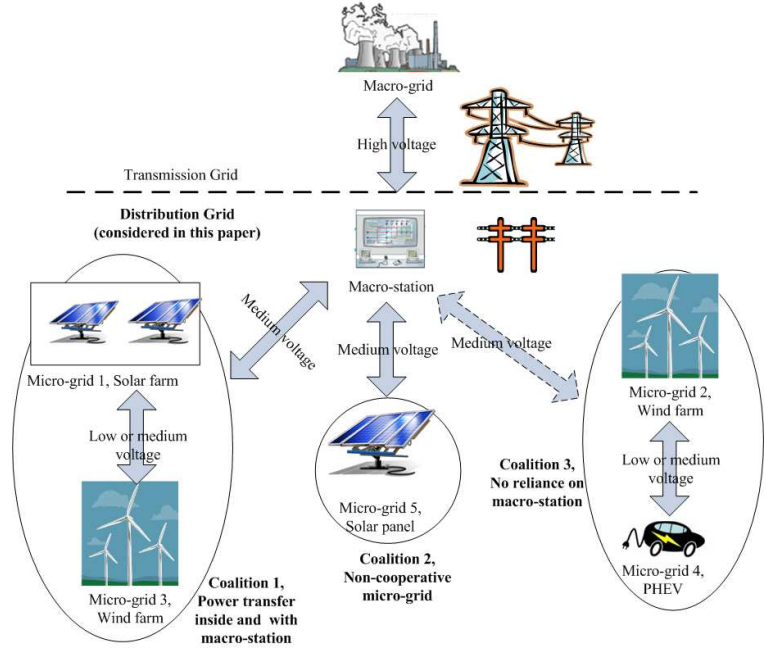}
    \vspace{-0.3cm}
    \caption{\label{fig:SysModel} An illustration of a cooperative micro-grid model.}
  \end{center}\vspace{-1cm}
\end{figure}

Instead of exchanging power exclusively with the substation, the micro-grids may want to form a cooperative group, i.e., a coalition, which constitutes a local energy exchange market. Inside this coalition, the micro-grids can transfer power locally among each other thus reducing the power losses and improving the autonomy of the micro-grid network. This reduction of wasted power is mainly due to two factors: (i)- several micro-grids may be closely located and, thus, can transfer power over shorter distances and (ii)- local power exchange can help to avoid the power losses at the level of the substation's transformer. Therefore, depending on their location and their power needs, the micro-grids have an incentive and a mutual benefit to cooperate so as to trade power locally within their given distribution network. An illustration of the proposed system model with cooperative coalitions is shown in Fig.~\ref{fig:SysModel} for $5$ micro-grids.  Note that, while Fig.~\ref{fig:SysModel} shows that the micro-grid elements are cooperating, this cooperation may be executed either using intelligent software agents or via an external entity that owns or operates the micro-grids.

\subsubsection{Game Theoretic Formulation and Results}
To study a cooperative energy exchange model such as in Fig.~\ref{fig:SysModel}, a coalitional game can be formulated between the micro-grids in the set $\mathcal{N}$. A coalition $S\subseteq \mathcal{N}$ is defined as a number of cooperative micro-grids which can be divided into two groups: the group of sellers denoted by $S_s \subset S$ and the group of buyers which we denote by $S_b \subset S$, such that $S_s \cup S_b = S$. Inside each coalition, the sellers in $S_s$ may exchange power with the buyers in $S_b$ as well as with the substation (if required).

The utility of a coalition $S$ can be expressed as a function of the members of $S$ as well as of the way in which sellers and buyers are matched (i.e., which seller is providing energy to which buyer). Matching the sellers and the buyers is a challenging task on its own, which can also be addressed using game theoretic techniques as discussed later. Nonetheless, given a certain buyer-to-seller association $\Pi$ resulting from any matching algorithm between buyers and sellers inside a coalition $S$, the utility of $S$ can be written as:
\begin{equation}\label{eq:utili}
u(S,\Pi) = - \left( \sum_{i \in S_s, j \in S_b} w_iP_{ij}^{\textrm{loss}} + \sum_{i \in S_s}w_i P_{i0}^{\textrm{loss}} +\sum_{j \in S_b} w_j P_{j0}^{\textrm{loss}}\right),
\end{equation}
where $w_i$ is a pricing factor, $P_{i0}^{\textrm{loss}}$ and $P_{j0}^{\textrm{loss}}$  are, respectively, the power losses during the distribution of power (if any) between the sellers and buyers of $S$ and the main substation while $P_{ij}^{\textrm{loss}}$ represents the power lost over the distribution lines during the local power transfer, inside $S$, between a seller $i$ and a buyer $j$. These losses are, as mentioned before, function of various factors such as distance and distribution voltage. There are two key questions that need to be answered: (i)- for a given coalition $S$, how can one match the sellers to the buyers inside $S$ so as to optimize (\ref{eq:utili})? and (ii)- how can the micro-grids interact to form coalitions such as in Fig.~\ref{fig:SysModel} so as to minimize the power losses?

To address the first question, one can adopt advanced techniques from two main game theoretic branches: auction theory and matching games~\cite{GT01,MC01,AUC00,AUC01}. Auction theory is essentially an analytical framework used to study the interactions between a number of sellers, each of which has some commodity or good to sell (in this example, the commodity is power), and a number of buyers interested in obtaining the good so as to optimize their objective functions. The outcome of the auction is the price at which the trade takes place as well as the amount of good sold to each buyer. In the studied model, given that, inside a coalition $S$ (here, a coalition refers to a fixed grouping of seller and buyer micro-grids), multiple buyers and sellers can exist, one suitable framework to model the matching between buyers and sellers could be through the use of a \emph{double auction}. Note that, whether a micro-grid acts as a seller or a buyer is dependent on its current generation and demand state, as previously mentioned. Thus, within every coalition $S$, one can formulate a double auction game in which the players are the buyers and sellers inside $S$. The strategies of every member in $S$ correspond to the price at which it is willing to buy/sell energy and the quantity that it wishes to sell/buy. The objective of each player is to determine the optimal quantity and price at which it wants to trade so as to optimize its objective function. Using techniques such as those developed in \cite{AUC00,AUC01} or in our work in \cite{WS03}, one can determine the price at the equilibrium along with the quantities that are being traded, i.e., the matching of sellers to buyers. Thus, the outcome of this auction determines the prices and quantities traded inside $S$. Subsequently, the quantities can be used to determine the payoff of each player in $S$ as in (\ref{eq:utili}), and, hence, the stability or result of the second question, i.e., how can the micro-grids form the coalitions whose answer can be found using the framework of coalition formation games. Note that, although (\ref{eq:utili}) incorporates only the quantities (through the power over the distribution lines), it can also be extended to become a function of the trading price resulting from the auction. Thus, a complete solution of the cooperative energy exchange problem requires solving two correlated games: an auction or matching game inside the coalitions and a coalition formation game to build the coalitions.

The work in \cite{WS01} presented a solution for the second question while adopting a heuristic seller-to-buyer matching procedure. First, we note that (\ref{eq:utili}) can be seen as the costs that a coalition $S$ pays due to the wasted power. These costs can be either actual monetary losses or virtual losses imposed by the power grid operator so as to control the wasted power in the network. In both cases, one can assume that (\ref{eq:utili}) is a \emph{transferable utility} in the sense that, to evaluate the payoff of every micro-grid in $S$, one can divide (\ref{eq:utili}) between the members of $S$, in any arbitrary manner.

The solution to the coalition formation game proposed in \cite{WS01} can be described as follows. Using an underlying communication network, at any period of time, the micro-grids can exchange or signal their energy needs (e.g., using agents). Group of micro-grids that find it beneficial to cooperate by evaluating (\ref{eq:utili}) can decide to join together and form a single coalition $S$. In order to evaluate (\ref{eq:utili}) and decide on whether to form a coalition or not, the involved micro-grids must: (i)- agree on the procedure for matching sellers to buyers (e.g., using the heuristic of \cite{WS01}) and (ii)- compute their prospective payoff which is found through a mapping that associates with every utility such as in (\ref{eq:utili}) a vector $\boldsymbol{\phi}^S$ of payoffs where each element $\phi_i^S$ is the individual utility of micro-grid $i$ when it is part of coalition $S$. This mapping can represent certain fairness rules or other criteria set by the power grid operator. Subsequently, a group of micro-grids would cooperate and form a single, larger coalition if this formation increases the payoff $\phi_i$ (reduces the power losses) of at least one of the involved micro-grids without decreasing the payoff of any of the others. Similarly, a coalition of micro-grids can decide to split and divide itself into smaller coalitions if it is beneficial to do so.

\begin{figure}[t!]
  \begin{center}
   \vspace{-0.2cm}
    \includegraphics[width=8cm]{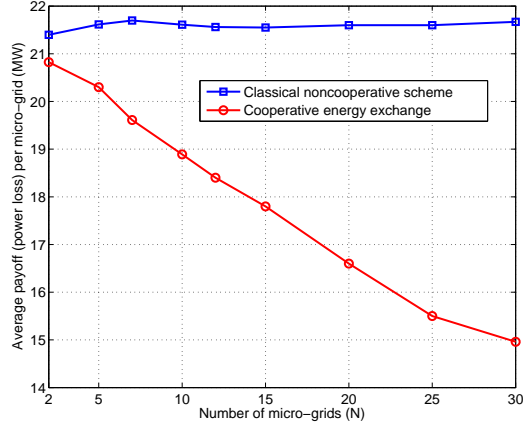}
    \vspace{-0.5cm}
    \caption{\label{fig:perf} Average power loss per micro-grid resulting from applying a cooperative game for energy exchange in micro-grid networks.}
  \end{center}\vspace{-1.1cm}
\end{figure}

Using this procedure for coalition formation, combined with a fixed heuristic for matching sellers to buyers, leads to promising results as shown using simulations in \cite{WS01}. For example, the results in Fig.~\ref{fig:perf} show that, compared to the classical noncooperative energy exchange scheme, the use of cooperative games yields a performance advantage, in terms of average power loss per micro-grid, which is increasing with the number of micro-grids $N$ and reaching up to $31\%$ of loss reduction (at $N=30$) relative  to the classical scheme. Further, the simulations in \cite{WS01} show which coalitions would emerge in a typical network as well as the overhead required for forming these coalitions.

Clearly, cooperative games could become a foundation for introducing local energy exchange between micro-grids in future smart grid systems. This local energy exchange could constitute one of the main steps towards the vision of an autonomous micro-grid network.

\subsubsection{Future Opportunities}
The model studied in \cite{WS01} can be used as a basis to develop more advanced and practical cooperative energy exchange models. In fact, several future opportunities for extending the work in \cite{WS01} can be explored:
\begin{itemize}
\item Proposing algorithms, based on auction theory or matching games, that can lead to optimal and stable associations between micro-grids that act as seller (i.e., have an excess of energy) and micro-grids that act as buyer (i.e., have a deficiency of energy).
\item Developing novel equilibrium concepts that are suitable to characterize a hybrid game composed of a coalition formation game for selecting the cooperative partners and an auction or matching game for solving the seller-to-buyer association problem.
\item Developing utility functions that capture, not only the power losses such as in \cite{WS01}, but also the prices during energy trade and the costs for communication overhead.
\item Studying dynamic cooperative game models that can capture the instantaneous changes in renewable energy generation and consumer loads.
\item Analyzing, using classical noncooperative games, the impact of storage on the outcome of the local energy exchange model as well as the coalitional game formulation.
\item Proposing a practical implementation that can be used as a testbed to enable cooperative energy exchange using game theory in future smart grid.\vspace{-0.2cm}
\end{itemize}

\subsection{Distributed Control of Micro-grids using Noncooperative Games}\label{sec:uiuc}\vspace{-0.2cm}
In classical large-scale power systems, any energy mismatch between demand and generation is often assumed to be compensated by a slack bus. However, in micro-grids, such a power reserve is often absent. Thus, it is of interest to have a mechanism which enables a distributed operation of the micro-grids, taking into account the individual constraints and objectives of each component. Moreover, most literature typically deals either with managing the load (demand-side) or the energy source (supply-side). In smart grids encompassing small-scale power components such as micro-grids, one must develop a generic framework that can capture both the competition over the energy resources that arises between the loads as well as the competition over the supply of energy that arises between the sources.

Toward this goal, the work in \cite{MGA02} proposes a noncooperative game approach for controlling both the loads and energy sources in a small-scale power system such as a micro-grid. Formally, the authors define a static noncooperative game in which the player set $\mathcal{N}=\mathcal{L} \cup \mathcal{S}$ represents the group of loads $\mathcal{L}$ and power sources $\mathcal{S}$ and the strategy of each player depends on its type. For a source node $i \in \mathcal{S}$, the strategy $a_i$ is chosen so as to regulate the voltage. The strategy space is typically the space of voltages that can be adopted, as explained in \cite{MGA02}. In contrast, for a load $l \in \mathcal{L}$, the strategy $a_l$ is chosen so as to control its variable shunt resistance to the ground. Often, for a load, the strategy space can be chosen as the set of possible values for the shunt resistance which is a closed subset of the space of real numbers~\cite{MGA02}.

The objective functions of the source and the load are application-dependent, but, in general, they will be function of the strategies, the currents, the voltages, and the impedance as discussed in \cite{MGA02}. For the source and load interaction game, the key question that must be answered pertains to how a load (source) can choose is strategy so as to optimize its objective function given the impact of this strategy on the source's (load's) strategic decision as well as on the strategies of the other loads (sources). The coupling between the strategies comes from the well-known dependence between the value of the resistance (strategy of the load) and the voltage (strategy of the source).

Under different objective functions and scenarios, the work in \cite{MGA02} studies the Nash equilibrium as a solution for this strategic noncooperative game between sources and loads in a small-scale power system.  On the one hand, for a simple one source, one load example in which the source wishes to regulate its terminal voltage while the load wishes to regulate its input power, the authors show that two Nash equilibria exist: (i)- an undesirable equilibrium which corresponds to the limiting case in which the maximum power transfer of the system is reached and the power regulation objective of the load cannot be fully met because of excessive line impedance and (ii)- a desirable equilibrium under which both load and source are able to regulate their power and optimize their utilities. For various other examples, the authors discuss several approaches for studying the existence and efficiency of the equilibria. Moreover, the use of Nash bargaining for improving the performance of a two-load game is also discussed in \cite{MGA02}. We note that, in many cases of interest, the work in \cite{MGA02}  provides algorithms for finding equilibrium solutions which are based on known algorithms such as those used in classical convex or mixed-integer optimization, for example.

Essentially, the work in \cite{MGA02} establishes how game theory can be used to define individual objectives for small-scale power systems such as micro-grids. This work clearly shows that, by allowing loads and sources to act in a distributed manner, one can have several insights on how these loads and sources can achieve their objectives (e.g., regulate their power). The main message that can be extracted from \cite{MGA02} is two-fold: (i)- the use of noncooperative games can adequately model the interactions between sources and loads in a small-scale power system and (ii)- advanced analytical techniques and algorithms are still needed to enable the operation at Nash equilibrium points as well as to improve the efficiency of these points. Moreover, one can envision several future directions that build upon \cite{MGA02}, as follows:
\begin{itemize}
\item Studying the impact of dynamics (e.g., variations in the sources generation rate) on the outcome of the game and proposing dynamic-game approaches to address this issue which is of central importance when the sources are renewable energy devices. This can be done by introducing notions of information and time evolution.
\item Developing algorithms that can characterize the equilibria for multi-player source/load games.
  \item Introducing additional players and strategies into the game such as enabling bus selection or introducing other power components as players.
\end{itemize}

Basically, game theory can constitute a solid foundation for enabling distributed control of loads and sources in small-scale power systems by developing individual objective optimization, distributed operation, and practical algorithms.\vspace{-0.2cm}

\subsection{Other Game-Theoretic Techniques in Micro-Grid Design}
Beyond cooperative energy exchange and distributed control, several other game theoretic applications in micro-grids can also be studied. For example, in \cite{MGA03}, the authors propose a game theoretic framework that enables the micro-grids to decide on whether to store or use energy so as to meet the predicted demand of their consumers. The essence of the framework is based on two types of games: a noncooperative solution for the Potluck problem and an auction game for determining the pricing in the micro-grid network.

The Potluck problem is essentially a formulation of the situation that involves two types of players: players that have a certain good to supply and players that have a certain need for this good. Essentially, the Potluck problem adopts noncooperative techniques to study how the players can decide, without communication, on the amount that they need to supply and the amount that they need to demand so as to reach a satisfactory equilibrium point in which the supply and the demand are equal. Reaching such a point may not possible if the players act rationally, i.e., try to improve their individual utilities~\cite{GT01}. In this case, the Potluck problem has no equilibrium and results in a system that oscillates between two states: one in which the demand exceeds the supply and one in which the supply exceeds the demand~\cite{MGA03}. To overcome this problem, the authors in \cite{MGA03} propose a learning scheme that enables a non-rational behavior of the players and which can reach a desired point of the system. Then, the authors complement their scheme with an auction algorithm that enables to study the pricing that emerges in the micro-grid energy exchange market. The results are focused on two-player games, but extensions to multi-player are also possible. Beyond \cite{MGA03}, wind turbine control, pricing issues, and cooperation in micro-grids are also discussed in \cite{MGA00,WS03,MGA01,MG04,MG05,MG06,WS01,MGA02}.

\subsection{Future Opportunities for Applying Game Theory in Micro-Grids}
\begin{table*}[!t] \vspace*{-0.72em}
\caption{Summary of game theoretic techniques in micro-grid distribution networks.}\vspace{-0.52cm}
\begin{center}
\begin{tabular}{|p{0.2\textwidth}|p{0.2\textwidth}|p{0.4\textwidth}|}
  \hline
  {\bf Application} &{\bf Game Theoretic Technique}& {\bf Main Future Extensions}\\

  \hline
Subsection~\ref{sec:coal}: Cooperative energy exchange between micro-grids (such as in~\cite{WS01}).& Coalitional games &
    \begin{itemize}
  \item   Use matching games or auctions for assigning sellers to buyers.
  \item Propose new equilibrium concepts for cooperative games with auctions.
   \item Include communication overhead and market prices.
   \item Study dynamic models and include storage capabilities.
      \end{itemize}
      \\

  \hline
 Subsection~\ref{sec:uiuc}: Distributed control of loads and sources in a small-scale power system (such as in~\cite{MGA02}).& Noncooperative Nash games &
    \begin{itemize}
  \item  Study the impact of variations in generation rates on the system.
  \item Develop algorithms for finding equilibria in multi-source, multi-load games.
   \item Study evolutionary game models that include notions of information and time.
   \item Develop heterogeneous games which comprise, beyond sources and loads, additional smart grid components as players with different strategies.
      \end{itemize}
      \\

  \hline
   Controlling the usage of stored micro-grid energy (such as in ~\cite{MGA03}).& Noncooperative Nash games, the Potluck problem, and auction theory&
    \begin{itemize}
  \item  Introduce a stochastic game model.
  \item Develop learning algorithms for multi-player storage control in micro-grids.
      \end{itemize}
      \\
        \hline
         \multicolumn{3}{|p{0.8\textwidth}|}{ Other future game theoretic applications in micro-grids could involve several types of games such as  facility location games, Stackelberg games, advanced Nash games, and others.} \\
        \hline
\end{tabular}\label{tab:micro}\vspace{-0.7cm}
\end{center}
\end{table*}
The autonomous nature of the micro-grids combined with the need for individual objectives implies that game theory can be one of the cornerstones of future micro-grids. Beyond the perviously presented applications, various future opportunities exist such as:
 \begin{itemize}
 \item Using facility location games for deploying micro-grids efficiently in a new electricity network as well as for locating aggregation stations for electric vehicles.
      \item Introducing noncooperative game models to enable an autonomous switching between the two main operating modes of the micro-grids: autonomous mode (island) and cooperative mode (in coordination with the main grid).
           \item Applying Stackelberg games to study the market and energy coordination between the main electricity grid and the micro-grid.
 \item Applying network formation games for enabling information coordination between micro-grids.
 \item Modeling the dynamics and interactions between electric vehicles and the grid using dynamic game theory.
 \end{itemize}
A summary of the different applications of game theory in micro-grids is shown in Table~\ref{tab:micro}.\vspace{-0.3cm}

\section{Demand-Side Management in Smart Grids}\label{sec:dsm}\vspace{-0.2cm}
In this section, we provide an overview on demand-side management in smart grids. Then, we study two main applications of game theory for demand-side management: scheduling of appliances and storage management. We conclude this section with an overview on potential future applications.\vspace{-0.2cm}
\subsection{What is Demand-Side Management?}\vspace{-0.2cm}
Demand-side management is an essential characteristic of current and future smart grid systems using which a utility company is able to control the energy consumption at the consumers' premises. For example, demand-side management at residential homes aims at reducing consumption by providing energy-efficient equipment and encouraging energy-aware consumption. Also, demand-side management can be used to provide incentives to the consumers to shift their consumption to hours during which the energy grid is less loaded (e.g., by providing lower prices during night hours).

Demand-side management often entails interactions between two main players: utility companies and consumers. Consumers can be residential houses, businesses, or even electric cars. These interactions can be at technical levels (i.e., interactions between smart meters and utility company control centers) as well as at social levels (i.e., service agreements between the utility companies and their consumers). Thus, implementing efficient demand-side management schemes involves a variety of challenges such as devising pricing schemes that enable efficient load shifting, implementing scheduling scheme for appliances, monitoring and shaping consumer behavior, among others.

Robust and smart demand-side management techniques are expected to lie at the heart of future power systems. In fact, enabling the interconnection of consumers, electric cars, micro-grids, and utility companies can only be made possible with efficient demand-side management techniques. The essence of demand-side management revolves around the interactions between various entities with specific objectives which are reminiscent of the players' interactions in game theory. In fact, game theory provides a plethora of tools that can be applied for pricing and incentive mechanisms, scheduling of appliances, and efficient interconnection of heterogeneous nodes.

Note that, here, we group together the two related areas of demand-side management and demand response models. Demand response models refer to the programs that utility companies use to encourage the grid users to dynamically change their electricity load (according to a certain signal from the company, such as pricing information)) so as to have short-term reduction in energy consumption. In essence, demand response models must be able to shape the demand or match the supply so as to better utilize the power system. In contrast, demand-side management refers to program that attempt to make the users more energy-efficient on a longer time-scale. Often, demand response models are included under the umbrella of demand-side management due to the close correlation between the two.

\subsection{Game Theory and Demand-Side Management}
Game theory has been extensively used for demand-side management and demand-response models in smart grids such as in~\cite{DR02,DR03,DR00,DR01,DR05,DR06,HAM00,MGA04}. For instance, the work in \cite{DR02} studies two market models suitable for matching the supply and shaping the demand in a smart grid system. The main focus is on providing markets in which the consumers can shed or increase their load (depending on whether there is a surplus or deficit of energy) so as to much the supply. The authors study oligopolistic markets and discuss the properties of the resulting competitive equilibria. The work in \cite{DR05} studies a more elaborate demand response model in which a time-varying pricing model is developed so as to align the objective function of each household appliance with the social welfare.

In \cite{DR03}, the authors apply a simple class of noncooperative games, the so-called \emph{congestion games}, as a means for performing dynamic pricing so as to control the power demand in an effort to achieve, not only net energy savings, but also an efficient utilization of the energy. The authors discuss the key characteristics of the demand side management congestion game and show how the equilibrium can be reached using a distributed algorithm. Several additional applications of demand-side management and demand-response models are found in \cite{DR00,DR01,DR06,HAM00,MGA04}.

In order to provide a better overview on how game theory can be applied for demand-side management, in this section, we start by analyzing a noncooperative game approach for modeling the interactions between a number of consumers and an energy generator or substation. Then, we discuss the use of noncooperative games for micro-storage management. We conclude with an outlook on future opportunities for game-theoretic demand-side management.\vspace{-0.2cm}

\subsection{Game Theory for Demand-Side Management through Energy Consumption Scheduling}\vspace{-0.2cm}\label{sec:dsm2}
\subsubsection{Introduction and Model} Classical demand-side management schemes such as direct load control and smart pricing are focused on the interactions between a utility company and each individual end-user. On the one hand, direct load control enables the utility company to control the appliances inside the home of each individual consumer, based on a certain agreement. On the other hand, the essence of smart pricing is to provide monetary incentives for the users to voluntarily shift their consumption and balance the load on the electricity grid. While these schemes have been extensively deployed, they are all focused on the individual user energy. However, the authors in \cite{HAM00} show that, instead of focusing only on the individual user consumption such as in classical schemes, it is better to develop a demand-side management approach that optimizes the properties of the aggregate load of the users. This is enabled by the deployment of communication technologies that allow the users to coordinate their energy usage, when this is beneficial~\cite{HAM00}.

Similar to \cite{HAM00}, we consider a power system with $N$ users and a single energy source, such as a substation. A wired or wireless technology interconnects the smart meters and the sources, hence, enabling them to communicate at any point in time. We let $\mathcal{N}$ denote the set of all users. Assuming time is slotted into hour-long intervals, at any given hour $h$ the total consumption of all users is denoted by $L_h=\sum_{i \in \mathcal{N}} l_i^h$, with $l_i^h$ being the energy consumption of user $i$ at hour $h$. This total consumption incurs a cost on the utility company which could reflect either a physical cost (i.e., costs for thermal generators) or a virtual cost that is used by the utility company so as to encourage an energy-aware behavior by the users~\cite{HAM00}. Practical cost functions such as thermal generation costs are increasing with the load and, often, strictly convex. As a result, let $\sum_{h=1}^H C_h(L_h)$ denote the total cost incurred on the utility company over a period of $H$ hours by all $N$ users with $C_h(\cdot)$ being a strictly convex and increasing function. Note that, for a certain load value, the cost function $C_h(\cdot)$ could lead to different costs depending on the hour during which this load is consumed.

Based on the cost $\sum_{h=1}^H C_h(L_h)$ the utility company would decide on how much to charge the users for the consumption during the $H$ hours. The dependence of the cost function on the total users' load $L_h$ implies that a change in the load of one user would impact the total cost of the utility company which, in turn, impacts the individual charges of the users. Hence, clearly, the users can be seen as independent decision makers whose choices of scheduling times and loads would impact one another. In this model, the objective is to enable the smart meters at the users premises to utilize automatic energy consumption schedulers so as to choose when to schedule appliances in order to minimize the total cost on the utility company and, subsequently, minimize the charges on each individual user. To address this problem, a game theoretic formulation is suitable as shown in \cite{HAM00} and discussed next.
\subsubsection{A Noncooperative Game for Scheduling Appliances}
Essentially, we are interested in devising a demand-side management scheme that enables to schedule the \emph{shiftable} appliances such as dish washers or dryers, while minimizing the overall energy consumption and, thus, the charges on the consumers. In this context, as in \cite{HAM00}, we can formulate a static noncooperative game in which the set of users $\mathcal{N}$ represents the players with the strategy of every player $i\in \mathcal{N}$ being a vector $\boldsymbol{x}^i$ which is formed by stacking up energy consumption schedule vectors of the form $\boldsymbol{x}_{i,a}=[x_{i,a}^1,\ldots,x_{i,a}^H]$ where $x_{i,a}^h$ is the energy consumption scheduled for an appliance $a$ by user $i$.

In this noncooperative game, each user $i$ needs to select its vector $\boldsymbol{x}_i$ so as to optimize a utility function $u_i(\boldsymbol{x}_i,\boldsymbol{x}_{-i})$ which is mainly a function of the cost function $C_h$ at each time $h$. The exact expression of the utility depends on how the utility company performs the billing as well as on the type and energy requirement of the users' appliances. Exact expressions for this utility were derived in \cite{HAM00} under the assumption that each user is billed proportionally to its total consumption. In consequence, we have a static noncooperative game which we refer to as the \emph{appliances scheduling game} and we can make several remarks on the properties of this game based on the results in \cite{HAM00}, as follows:
\begin{enumerate}
\item A Nash equilibrium for the appliances scheduling game always exists and all equilibria coincide with the optimal scheduling policy that \emph{minimizes} the overall utility company cost which is given by $\sum_{h=1}^H C_h(L_h)$.
\item The Nash equilibrium of the game corresponds to a \emph{unique} set of total loads $l_i^{h,\textrm{NE}}$ at each user $i \in \mathcal{N}$.
\item Each user can map the total load $l_i^{h,\textrm{NE}}$ at the equilibrium to \emph{any} feasible set of strategies $\boldsymbol{x}_i^\textrm{NE}$. For the utility function considered in \cite{HAM00}, the appliances are, thus, indifferent to when they are scheduled as every schedule would always correspond to the minimum of the total cost incurred on the utility company.
\end{enumerate}

 The authors in \cite{HAM00} propose an algorithm that uses best response dynamics to find the Nash equilibrium while ensuring that no user has an incentive to cheat and announce an incorrect energy schedule. A \emph{best response} algorithm mainly relies on a sequence of decisions in which each player chooses the strategy that \emph{maximizes} its utility, given the current strategies of the other players. It is shown in \cite{HAM00} that, for the appliances scheduling game, best response dynamics always converges to an equilibrium. The simulations in \cite{HAM00} also show that, whenever consumers have a good number of shiftable appliances, adopting a game-theoretic approach for scheduling these appliances can reduce the energy costs of up to $18\%$ compared to existing solutions while also reducing the peak-to-average ratio of the energy demand (i.e., the ratio of the energy at peak hour to the average energy over a time period $H$) of about $17\%$.

\subsubsection{Future Extensions}
Clearly, using noncooperative games can lead to smarter demand-side management schemes. The model studied in this subsection can be extended in a variety of ways such as by:
\begin{itemize}
\item Introducing a utility function in which the time at which an appliance is scheduled impacts the payoff of the users. The objective of the game becomes to optimize a tradeoff between minimizing the charges and optimizing the appliances' waiting time. By doing so, the properties and results of the game formulated in \cite{HAM00} are significantly impacted, although the noncooperative framework is still useful to analyze the problem.
\item Considering multiple energy sources and the interactions among them. In such a setting, hierarchical games such as Stackelberg games are a good candidate to provide insights on the appliances' scheduling problem.
\item Studying a stochastic game counterpart of this model in which the smart meters schedule the appliances instantaneously based on the time-varying conditions of the network (e.g., the varying generation conditions of the energy source). The studied game can, in fact, constitute a building block for such a stochastic formulation. For instance, a stochastic game is essentially a dynamic game composed of a number of stages and in which, at the beginning of each stage, the game is in a specific state. In such a setting, the studied game and its solution can be used to solve or study each one of these stages. Hence, the studied game can serve as a single stage in a stochastic game setting (under both complete and incomplete information).
\end{itemize}

Each one of these extensions leads to new challenges but also contributes to the deployment of smart demand-side management schemes that account for the aggregate user loads as well as the individual objectives of the users.\vspace{-0.3cm}
\subsection{Demand-Side Management with Storage Devices}\label{sec:storev}\vspace{-0.2cm}
\subsubsection{Introduction and Model}
In the previous subsection, we focused on how the users can schedule their appliances so as to minimize their billing charges. The underlying assumption was that the users are acquiring energy so as to immediately use it for their appliances. However, in future smart grids, energy storage is expected to be a key component in smart homes, and, thus, it has a strong impact on demand-side management. For example, a user may decide to store energy during off-peak hours and use this stored energy to schedule its appliances, instead of obtaining this energy directly from the substation during peak hours. The introduction of smart home storage systems could significantly improve the energy-efficiency of the electricity market, but is also accompanied with several challenges. For example, if all homes decide to charge their energy storage devices concurrently, this could lead to an excess in demand which can eventually be detrimental to the overall system. In addition, the storage behavior of the users could impact the pricing in the market, and, thus, some users might find it better off to buy from the market than use a storage device (considering factors such as the cost and lifespan of such a device).

Here, we are interested in a demand-side management model such as the one introduced in \cite{MGA04} in which the users strategically choose on how to use their storage devices and when to buy energy. Hence, it is of interest to enable strategic storage decisions by the consumers during demand-side management, so as to optimize a variety of factors. In this respect, consider a set $\mathcal{N}$ of consumers each of which has a certain load $l_i^h, i \in \mathcal{N}$ at any given time interval $h$. At each interval $h$, every user $i \in \mathcal{N}$ can decide on its storage profile $s_i^h$ which corresponds to the amount of energy that $i$ is willing to charge and/or discharge. In this setting, as discussed in \cite{MGA04}, the objective is to enable the users to choose their storage profile $s_i^h$ at every hour $h$ so as to minimize the energy cost incurred over a whole period $H$, i.e., a whole day for example. Clearly, this strategic setting requires a game theoretic formulation, as seen next.

\subsubsection{Noncooperative Game Formulation and Results}
To determine the storage choices of the consumers, a noncooperative game can be formulated in which the players are the users in $\mathcal{N}$, the strategies are the storage profiles $s_i^h$ chosen at every time $h$, and the utilities are the costs incurred on the users over the whole day. In other words, in this noncooperative game, the objective of each user is to decide, over a period $H$, on a storage profile $\boldsymbol{s}_i =[s_i^1,\ldots,s_i^H]$ so as to optimize the following utility function~\cite{MGA04}:
\begin{equation}\label{eq:ag}
u(\boldsymbol{s}_i,\boldsymbol{s}_{-i}) = - p_i(\boldsymbol{s}_i,\boldsymbol{s}_{-i}) \sum_{h = 1}^H (s_i^h + l_i^h),
\end{equation}
where $\boldsymbol{s}_{-i}$ is the vector reflecting the storage profiles of all players except $i$ and $ p_i(\boldsymbol{s}_i,\boldsymbol{s}_{-i})$ is the price in the energy market which can be determined using approaches such as auctions or supply curves. Hereinafter, we consider that the price is determined from a continuous and increasing supply curve such as in \cite{MGA04}.

We note that, for any user $i \in \mathcal{N}$, the \emph{feasible} storage profiles $s_i^h$ that optimize (\ref{eq:ag}) is subject to several constraints that depend on three main characteristics of the storage device as described in \cite{MGA04}: its maximum capacity $e_i$, its storage efficiency $\alpha_i$ which reflects the fraction of energy that can be extracted after storage, and its running cost $c_i$ which depends on the physical characteristic of the storage device. The work in \cite{MGA04} shows that, under the considered constraints, for storage devices with homogeneous characteristics, the Nash equilibria of the game correspond to the storage profiles that minimize the global generator costs which are given by $\sum_{h=1}^H\int_0^{q_h}b_h(x)dx$, where $b_h(\cdot)$ is the supply curve and $q_h$ is the total amount of energy traded by \emph{all users} at time $h$ (including storage and load demand).

Using these results, the work in \cite{MGA04} studies the proposed game under two scenarios: an ideal scenario in which the players have a complete information about the market throughout time and an adaptive scenario in which the users adapt their strategies using a day-ahead best response mechanism. In the latter scenario, the essence of the algorithm is to allow the users to update their strategies based on their day-ahead knowledge of the market. These users can, then, adapt their strategy continuously using their predictions on the market trends. The noncooperative game devised in \cite{MGA04} is tested on empirical data from the UK market. The simulation results show that the learning scheme converges to a Nash equilibrium while reducing the peak demand which also leads to reduce costs and carbon emissions. The results also discuss the benefit of storage and how it impacts the social welfare of the system. Hence, these results show that the use of game theory has, not only theoretical impacts, but also practical significance in future smart grids.

\subsubsection{Future Extensions}
Storage devices are expected to proliferate in most future smart grid systems. The work in \cite{MGA04} showed that game theory can provide interesting solutions for managing storage at the users' side, i.e., within the context of demand-side management. To this end, we can foresee several future extensions for this area:
\begin{itemize}
\item Optimizing jointly the scheduling and storage profiles. For instance, in the studied storage game, the focus was solely on optimizing the storage profile (i.e., whether to discharge or charge the storage device at the customer premises), under a certain scheduled load. However, as seen in the game of Section~\ref{sec:dsm2}, one can also optimize the scheduling of appliances. Naturally, the scheduling and storage games are inter-related, since the result of the scheduling impacts how the customer uses its storage device and vice versa. Hence, one important future direction is to integrate both games into a joint scheduling and storage game in which the utility in (4) is optimized, not only with respect to the storage profile, but also with respect to the loads which can  be determined by using a modified version of the scheduling-only game in Section~\ref{sec:dsm2}.
\item Considering that the energy source is also strategic and aims at maximizing its revenue. Here, one must include an additional player in the game whose objective might not be aligned with the individual consumers' goals.
\item Allowing the consumers not only to use their storage devices but to also use privately owned renewable energy sources or solar panels. In such a setting, each consumer has three options (use storage, buy from market, generate) instead of only two (use storage, buy from market).
\item Developing learning algorithms that do not rely solely on the day-ahead predictions such as in \cite{MGA04}. To do so, one can adopt techniques from stochastic games to model the demand-side management problem with storage.
\item Studying advanced techniques for generating the prices and their impact on the storage profiles.\vspace{-0.3cm}
\end{itemize}

\subsection{Future Game Theoretic Approaches for Demand-Side Management}\vspace{-0.1cm}
 One of the key challenges of the future smart grid is designing demand-side management models that enable efficient management of the power supply and demand. Demand-side management schemes will always face technical challenges such as pricing, regulations, adaptive decision making, users' interactions, and dynamic operation. All of these issues are cornerstones to game theory, and, hence, this area is ripe for game theoretic techniques. In fact, demand-side management is perhaps the most natural setting for applying game theory due to the need of combining economical aspects such as pricing with strategic decision making by the various involved entities such as the suppliers and the consumers. Beyond the application described so far, several potential demand-side management games can be studied:
\begin{table*}[!t] \vspace*{-0.72em}
\caption{Summary of game theoretic techniques for demand-side management.}\vspace{-0.52cm}
\begin{center}
\begin{tabular}{|p{0.2\textwidth}|p{0.2\textwidth}|p{0.4\textwidth}|}
\hline
  {\bf Application} &{\bf Game Theoretic Technique}& {\bf Main Future Extensions}\\
  \hline
Subsection~\ref{sec:dsm2}: Energy consumption regulation through appliances' scheduling (such as in~\cite{HAM00}).& Noncooperative Nash games &
    \begin{itemize}
  \item  Develop noncooperative games with not only multiple consumers, but also multiple energy sources and utility companies.
  \item Enable the consumers to optimize the tradeoff between waiting time and billing charges for shiftable appliances.
   \item Study the dynamic game counterpart of the model.
      \end{itemize}
      \\
  \hline
Subsection~\ref{sec:storev}: Demand-side management with storage (such as in~\cite{MGA04}).& Noncooperative Nash games &
    \begin{itemize}
  \item  Consider a strategic energy source whose objective is not aligned with that of the consumers.
  \item Introduce the notion of privately owned micro-grids which impact the storage and usage strategies of the consumers.
   \item Develop intelligent equilibrium learning algorithms that account for instantaneous changes in the power system parameters.
      \item Develop joint appliances scheduling and storage optimization using stochastic games.
      \end{itemize}
      \\
        \hline
Demand response market models (such as in \cite{DR02,DR03,DR00,DR01,DR05}).& Auction games and/or noncooperative games &
    \begin{itemize}
  \item Develop strategy-proof techniques for demand response markets that are robust to cheating.
  \item Develop demand response models suitable for intermittent energy sources such as wind turbines.
    \item Study more elaborate differential game models for dynamic pricing in demand-response markets.
      \end{itemize}
      \\
              \hline
         \multicolumn{3}{|p{0.8\textwidth}|}{ Other future applications in demand-side management could involve several game-theoretic techniques such as  stochastic games, online learning techniques, and Bayesian games.} \\
        \hline
        \end{tabular}\label{tab:dsm}\vspace{-0.7cm}
\end{center}
\end{table*}
\begin{itemize}
\item Developing online algorithms for learning the Nash equilibria of demand-side management noncooperative games that involve short-term optimization of demand to match the supply, i.e., demand-response models. These algorithms can be based on stochastic games.
\item Studying the use of cooperative games for enabling a coordinated load management among the users which can, subsequently, lead to a more efficient load distribution and less costs on the utility operator. Such a model can be an extension of the coalition formation game presented for the micro-grids in Subsection~\ref{sec:coal} to the demand-side of the power grid (i.e., applied at user-level instead of micro-grid or smart grid level).
\item Studying the application of Bayesian games (i.e., games with imperfect and incomplete information) to develop noncooperative techniques that the consumers can use when little is known about other consumers' behavior.
\item Investigating the impact of privacy on the demand-side management games.
\end{itemize}
A summary of the different applications of game theory for demand-side management is shown in Table~\ref{tab:dsm}.\vspace{-0.3cm}

 \section{Communication in the Smart Grid}\label{sec:com}\vspace{-0.2cm}
 In this section, we survey the challenges of integrating communication technologies in smart grids and we discuss a network formation game for multi-hop communications. Then, we discuss future opportunities for game theory in smart grid communication.\vspace{-0.3cm}
 \subsection{Communication Technologies in the Smart Grid}\vspace{-0.2cm}
 One of the main characteristics of the smart grid is the ability to ensure a reliable information flow between a number of heterogeneous nodes~\cite{SG01}. On the one hand, the smart grid elements must be able to communicate information such as outage management to the utility company's control center~\cite{SG01}. On the other hand, smart meters need to communicate with nearby control centers so as to exchange information such as meter readings, pricing, or other control data~\cite{AS02}. Moreover, the communication of a load signal from the utility operator to electric vehicles or PHEVs is expected to be an inherent component of smart grid communications~\cite{PHEV00,PHEV01,UC02}. More recently, enabling two-way communications between the grid and PHEVs (or electric vehicles) has also received considerable attention in the research community~\cite{PHEV08,PHEV03,PHEV04,PHEV06,PHEV09}. In addition to these contributions, the Pacific Gas and Electric Company~(PG\&E) teamed up with Google and other car vendors to implement and showcase some potential applications of the vehicle-to-grid technology~\cite{PHEV07}. Furthermore, both the Electric Power Research Institute~(EPRI) and the National Institute of Standards and Technology~(NIST) have published several use cases in which a two-way communication between the grid and the vehicles is required such as for diagnostics~\cite{UC00}, vehicle roaming~\cite{UC01}, or others~\cite{UC04}. Hence,  even though electric vehicles currently rely mainly on the grid to vehicle communication, it is important to note that various research directions are being conducted to better understand the potential of vehicle-to-grid communications and its challenges~\cite{PHEV00,PHEV01,UC02,PHEV08,PHEV03,PHEV04,PHEV06,PHEV09,PHEV07,UC00,UC01,UC04}. Clearly, enabling many of the smart grid applications discussed so far such as demand-side management or micro-grid coordination as well as the integration of new elements such as electric vehicles is contingent upon the deployment of an efficient and reliable communication architecture that can truly allow a ``smart'' operation of future power systems. 

 Several technologies such as power line communication~(PLC)~\cite{AS00,AS01,AS02,AS03,AS07,AS08,WSPLC01} or wireless communication~\cite{WI00,WI01,WI02,WI03} are being considered as candidate communication techniques for the smart grid. Each technology has its advantages and shortcomings as well as its own range of applications. For instance, PLC is being considered for deployment by utility companies around the world for smart metering, load control, or other applications~\cite{AS01,AS02,AS03,AS07,AS08}. However, the use of PLC for high-data rate applications is still in its infancy. Many advocates of PLC argue that it can be one of the main communication technologies in the smart grid due to the low cost of its deployment~\cite{AS01,AS02,AS03,AS07,AS08}. In contrast, several recent works have proposed wireless communication techniques such as cognitive radio to serve as the main networking technology in the smart grid~\cite{WI00,WI01,WI02,WI03}.

In practice, the future smart grid will consist of a myriad of communication technologies that must coexist and operate efficiently together~\cite{SG01,SG000}. The communication layer of the smart grid will most likely be composed of both wireless and wireline technologies~\cite{AS02}. Deploying the right technology for the right application is still an open problem~\cite{SG01}. For example, one can leverage the already existing power lines in the distribution network and adopt PLC for advanced metering infrastructures, however, to do so, many challenges such as reliability and performance need to be addressed. In contrast, for long range transmission, wireless techniques may provide a better alternative due to the possibility of adopting advanced approaches such as cognitive radio or cooperative communications~\cite{WI03}. Also, security and privacy considerations might be a central issue in deciding on the best communication technology for the future grid~\cite{AS02}.

The integration of communication networks into a large-scale system such as the smart grid increases the complexity in network design and analysis motivating the use of advanced tools such as game theory due to its proven efficiency in wireless and wireline communications~\cite{MC01}. In this section, first, we investigate how network formation games can be used to allow the smart grid elements to interact and perform multi-hop narrowband PLC communication. Then, we discuss other existing opportunities for developing game theoretic frameworks tailored to smart grid communications.\vspace{-0.3cm}


 \subsection{Game Theory for Multi-hop Power Line Communications}\label{sec:mhop}\vspace{-0.1cm}
 \subsubsection{Introduction and Model}
Within large-scale networks such as the smart grid, PLC is one of the candidate technologies that can be used to ensure data communication between the different smart grid elements such as sensors that are typically used to collect data (e.g., household loads, monitoring data, maintenance, prices inquiry, etc.) and transmit it to a control node~\cite{AS00}. Enabling such PLC-based applications faces a variety of challenges. For instance, one main weakness of PLC is its reliability in communicating important smart grid data such as outage management information~\cite{PLC00}. This challenge stems from the fact that, in a PLC system, a line outage affects the entire set of devices down stream from that line failure.  Those down stream devices will be unable to report their status to this grid. Hence, when designing PLC systems for the future grid, one must incorporate such reliability issues in the overall design, using either backup lines or alternative line failure detection techniques such as in \cite{RC02}. We do note that, the reliability problems of PLC are one of the main motivations for research works which investigates alternative communication techniques such as wired communication networks or wireless technologies (e.g., ZigBee, cognitive radio, etc.). Beyond reliability, PLC design also faces additional technical challenges such as channel modeling, medium access, efficient data transmission, and advanced network planning.

Despite these challenges, the use of narrowband PLC, which is a version of PLC which operates on narrowband frequencies, constitutes a strong candidate for smart grid applications and has already been widely adopted for deploying advanced metering infrastructure in Europe~\cite{AS01,AS02,AS03,AS07,AS08}. One of the main impediments of narrowband PLC is its limited channel capacity which is shown to decrease rapidly with the communication distance as discussed in \cite[Chap.~5]{PLC00}. As a result, developing intelligent and advanced algorithms that can overcome these capacity limitations in smart grid systems is of central interest and a key enabler for several interesting smart grid applications.

Let us consider the set $\mathcal{M}$ of $M$ physically interconnected smart grid elements. Here, a smart grid element could represent any type of components in the smart grid that is equipped with communication capabilities such as smart sensors or other entities.  Each smart element in $\mathcal{M}$ needs to communicate different information such as control data, load reports, pricing inquiries, or event detection data to a common access point~(CAP). This access point can be either a control center installed by the grid operator or a repeater that connects the considered area to other parts of the smart grid. This model is based on our work in \cite{WSPLC01}. This communication can be enabled by using narrowband PLC operating over a frequency range between $3$~kHz and $500$~kHz.\footnote{Note that, in Europe, the maximum frequency for narrowband PLC is $148.5$~kHz while in the USA it can go up to $500$~kHz~\cite{PLC00}.} The smart elements typically transmit their data or control information directly to the CAP. During this process, the capacity $C_{i,\textrm{CAP}}^k$of any point-to-point PLC communication link between a smart element $i \in \mathcal{M}$ and the CAP, using a certain frequency $k$ (assigned by the CAP) is given by the so-called \emph{water-filling} solution~\cite[Chap.~5]{PLC00}:
\begin{equation}\label{eq:capacity}
C_{i,\textrm{CAP}}^k = \int_{f \in \mathcal{F}^B_{i,k}}1/2 \log_2\left[\frac{B}{N(f)}\right]df,
\end{equation}
where  $N(f)$ is a colored background noise and $ \mathcal{F}^B_{i,k}$ is the range of frequencies for channel $k$ (which depends on the bandwidth) for which we have $N(f)\le B$, where $B$ is the solution to
\begin{equation}\label{eq:capacity3}
P_{i,\textrm{re}}=P_i\cdot10^{-\kappa d_{i,\textrm{CAP}}}= \int_{f \in \mathcal{F}^B_{i,k}}\left[B-N(f)\right],
\end{equation}
with $P_i$ being the transmit power of the smart element, $\kappa$ being the attenuation factor which ranges between $40$~dB/km and $100$~dB/km, and $d_{i,\textrm{CAP}}$ being the distance between $i$ and the CAP.

It is known that the capacity in (\ref{eq:capacity}) is large for small distances, however, it can decay very fast with distance~\cite{PLC00}. This decrease in capacity can lead to an increase in the delay during the communication between the smart elements and the CAP. Many of the emerging applications within smart grid networks such as demand-side management can require a near real-time communication delay and, thus, high capacities. In consequence, it is of interest to design an improved architecture that enables the smart elements to utilize narrowband PLC for sending their data, while maintaining reasonable delays.

While most of the smart elements (such as smart meters) that are currently being deployed communicate exclusively with the control center, the need for advanced sensing and data collection in the smart grid has incited many advanced communication architectures in which the smart grid elements communicate, not only with the utility company or the control center, but also with one another (e.g., by forming mesh or multi-hop architectures) as discussed in~\cite{PLC00,AS02,AS07,SM00,SM01,SM02,SM03,SM05,SM06,QZ01}. To this end, one possible way to overcome the limited capacity of PLC communication is to enable a multi-hop PLC architecture as discussed in \cite{WSPLC01,PLC00,SM01,AS09} by leveraging on the possibility of having smart elements which can use PLC to communicate, not only with the control center but also with one another as in~\cite{PLC00,AS02,AS07,SM00,SM01,SM02,SM03,SM05,SM06,QZ01}. By allowing the smart elements to relay each other's data, the transmission delay can be reduced due to two key characteristics: (i)- the capacity in (\ref{eq:capacity}) is very large at small to medium distances and (ii)- several groups of smart elements may be physically co-located or neighboring (e.g., nearby homes) and, if well-equipped, they can communicate with one another. To perform this multi-hop PLC communication one of the main questions that needs to be answered is ``how can the smart elements interact and pick their next hop so as to optimize their delay?'' As seen next, game theory can provide us with an answer to this question.

 \subsubsection{Network Formation Game: Formulation and Results}
To improve its communication delay, each smart element may be able to communicate with a higher PLC capacity by using multi-hop transmissions. Thus, each smart element needs to decide on the \emph{path} that it will use to reach the CAP which can be either a direct communication path or a multi-hop path. The smart element and the CAP would eventually be connected using a logically defined \emph{directed} network graph $G(\mathcal{M},\mathcal{E})$ with $\mathcal{M}$ being the set of vertices of the graph and $\mathcal{E}$ being the set of all edges (links) between pairs of smart elements. Our objective is to study how the smart elements can form this network graph, strategically.

To do so, we use the framework of network formation games, as done in \cite{WSPLC01}. Network formation games involve situations in which a number of players need to interact in order to decide on the formation of a network graph among them~\cite{MC01}. For multi-hop PLC, the network formation game is defined among the smart elements in $\mathcal{M}$ who seek to communicate with the CAP over a communication tree structure that is rooted at the CAP. Note that, in this model, it is assumed that all smart elements are physically interconnected such that PLC is possible between any pair. Certainly, if some smart elements cannot reach each other using PLC, then they are not involved in the game. In this game, the objective of each smart element $i \in \mathcal{M}$ is to select the path that minimizes its overall transmission delay when sending its data to the CAP. Hence, given any tree structure $G$  resulting from the strategy selections of all the smart elements in $\mathcal{M}$, the cost function of any $i \in \mathcal{M}$ in the current graph $G$ can be expressed by~\cite{WSPLC01}:
\begin{equation}\label{eq:delnet}
c_i(G)=  \sum_{(i_l,i_{l+1}) \in q_i} \tau_{i_l,i_{l+1}}.
\end{equation}
Here, $q_i=\{i_1,\ldots,i_L\}$, with $i_1 = i$ and $i_L$ being the CAP, represents the multi-hop path from $i$ to the CAP and $\tau_{i_l,i_{l+1}}$ is the delay experienced during the transmission from smart element $i_l$ to smart element $i_{l+1}$ which can be given by:
$\tau_{i_l,i_{l+1}} = \frac{R\cdot L_i}{ C_{i_l,i_{l+1}}^k},$
where $L_i$ is the number of packets of $R$  bits that $i$ needs to transmit and $C_{i_l,i_{l+1}}^k$ is the capacity for the narrowband PLC transmission between $i_l$ and $i_{l+1}$ over channel $k$. In this model, similar to \cite{WSPLC01}, it is assumed that there exists a pre-determined frequency allocation scheme that selects the frequencies used between each two pairs of smart elements. Moreover, each smart element can only accept a limited number of connections due to the limited amount of available frequencies and bandwidth that it can allocate.

In this game, the objective of each smart element is to find its preferred partner for forwarding the packets and each smart element is assumed to act on its own, without coordinating its strategy with neighboring meters. In other words, if, at some point in time, a smart element $1$ finds it beneficial to connect to a smart element $2$, it will not coordinate this choice with any of the other smart elements who may be impacted (positively or negatively) by smart element $1$'s choice. Although a network formation game involves some form of cooperation (i.e., the smart elements helping each other), the lack of coordination during the decision making process implies that a noncooperative solution model is more suitable. Certainly, a fully cooperative game counter-part of this model in which all smart elements jointly coordinate their strategies (similar to the joint coordination for energy management done in Section~\ref{sec:coal}) can also be studied in future work in this area and its implications in terms of improved performance and additional costs in terms of information exchange can be analyzed.

Accordingly, to form the network graph, an algorithm based on noncooperative techniques can be developed as shown in \cite{WSPLC01}. Essentially, one can consider a noncooperative game in which the players are the smart elements with the strategies being discrete sets representing the choices of a next hop done by the players. The smart elements can choose their multi-hop routes by using best response dynamics in which each smart element chooses the link that minimizes its delay in (\ref{eq:delnet}), given an observed set of strategies from the other players, i.e., the current graph $G$. This process, as discussed in \cite{WSPLC01}, leads to a Nash network, i.e., a network in which no smart element can improve its delay by changing its chosen path.

\begin{figure}[t]
\begin{center}
\includegraphics[angle=0,width=8cm]{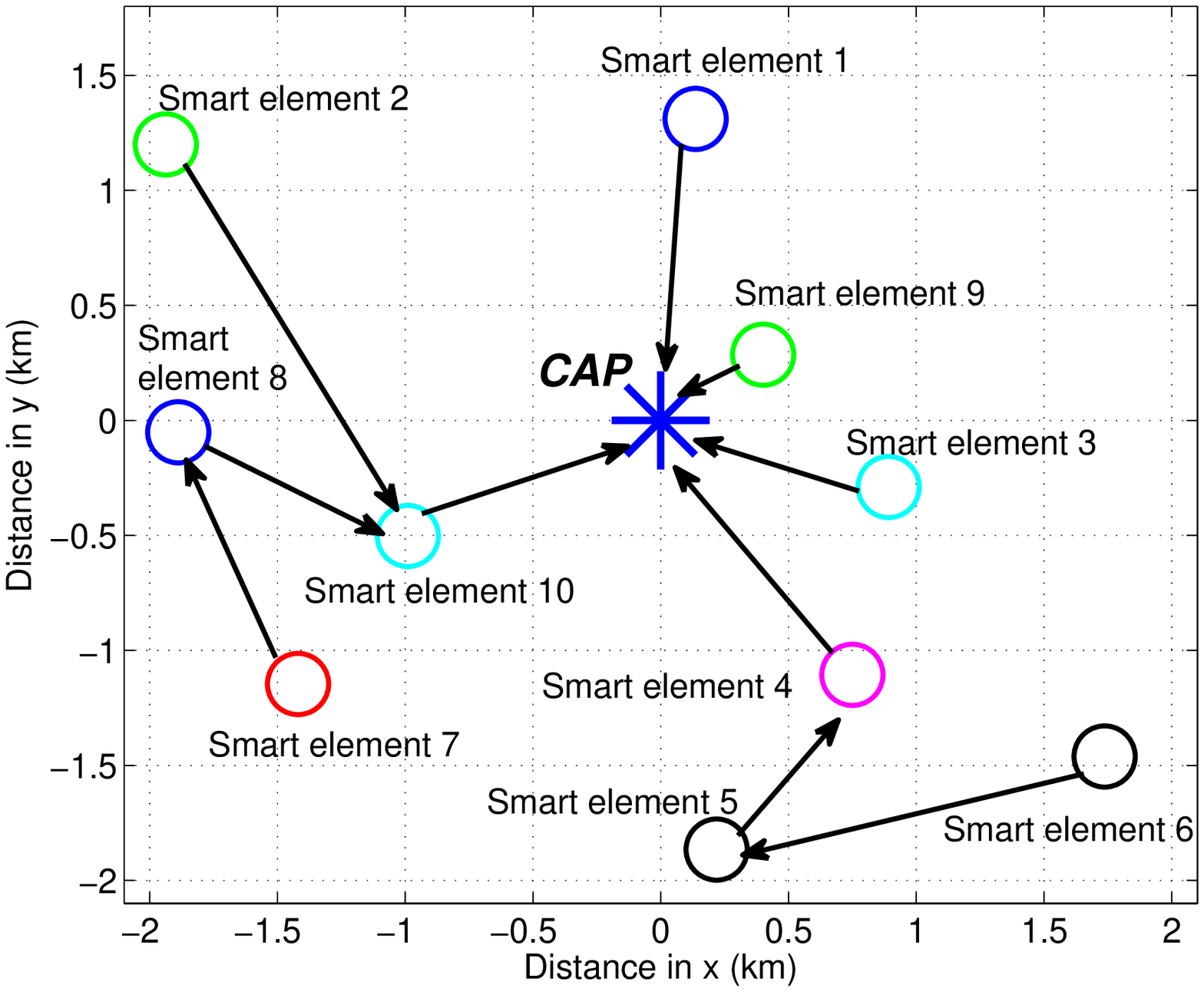}
\end{center}\vspace{-0.7cm}
\caption {A Nash network interconnecting $M=10$~randomly deployed smart elements that interacted using a network formation game.} \label{snapshot}\vspace{-0.8cm}
\end{figure}

In Fig.~\ref{snapshot}, based on \cite{WSPLC01}, we show a sample network consisting of a tree structure resulting from a network formation game in a system with $M=10$~randomly deployed smart elements. Fig.~\ref{snapshot} shows that, while smart elements close to the CAP such as $3$ and $9$, prefer to communicate directly with the CAP, other far away elements such as $6$ or $7$ obtain a better delay by using a two-hop link instead of a direct CAP connection. This demonstrates that adopting a multi-hop network formation game has a strong potential of improving the delay during PLC communication, if communication between the smart elements is enabled. The network in Fig.~\ref{snapshot} is a Nash network as no smart element can decrease its delay by unilaterally changing its current link. For example, consider smart element $6$ whose feasible strategies are all other smart elements and the CAP. If smart element $6$ decides to disconnect from $5$ and connect to smart elements $1$, $2$, $7$, $8$, $9$, or $10$ its delay increase from $102.5$~ms to about $382$~seconds. Alternatively, if it connects to smart element $3$, its delay increase from $10$~ms to $153$~ms and if it connects to smart element $4$, its delay increase from $102.5$~ms to $120.1$~ms. Hence, clearly, smart element $6$ has no incentive to change its current strategy. Moreover, the simulations in \cite{WSPLC01} have further shown that adopting network formation games for multi-hop PLC can reduce the average delay per smart element of at least $28.7\%$ and $60.2\%$  relative to the star network and a nearest neighbor algorithm, respectively.

\subsubsection{Future Extensions}
Using multi-hop PLC in future smart grid looks like a promising approach for enabling some interesting applications such as advanced metering. To this end, one can investigate several challenges of multi-hop PLC that go beyond the work done in \cite{WSPLC01}, most notably:
\begin{itemize}
\item Introducing a dynamic approach for jointly performing network formation and channel allocation in a PLC network.
\item Enabling the smart elements to make strategic decisions based, not only on their current observation of the network, but also based on long-term observation goals (i.e., using \emph{foresighted} network formation games).
\item Analyzing implementation and deployment issues for allowing multi-hop PLC using network formation games. To do so, one must investigate several practical issues such as interference, measurements, and others.
\end{itemize}

 \subsubsection{Future Opportunities for Game Theory in Smart Grid Communications}
Communication protocols for smart grid systems are still at their infancy. Most existing approaches are mainly focused on integration issues and projected implementations. In the next few years, the proliferation of novel services operating over the smart grid will certainly require advanced communication strategies. To this end, applying game theory for designing future communication protocols for the smart grid is a promising area with little existing work so far. Many of the foreseen game theoretical approaches will be tailored to the communication networking underlying the future smart grid services. For example, in Sections~\ref{sec:mg} and \ref{sec:dsm}, we developed several techniques for emerging applications in micro-grids and demand-side management. Most of these approaches assume that the communication network is reliable and already deployed. However, the nature of the communication network may strongly impact the efficiency of the application. As an example, a communication network with high delays can lead to wrong pricing inquiries and, hence, strongly affect the demand-side management algorithms. Hence, it is of interest to study how these game theoretical techniques can be improved so as to account for the constraints of the communication architecture being employed and its impact over these applications. Moreover, one important future direction is to enable the components of the smart grid to \emph{strategically} decide on the communication protocol that they will adopt, depending on their application constraints.

Beyond this, it is expected that a variety of wireless technologies need to co-exist in future smart grids. For example, short-range technologies such as ZigBee or Bluetooth may be used for in-home applications, while long-range wireless technologies such as cellular networks or classical wired technologies can be used for transmission over long distances~\cite{AS02}. In this context, game theory is expected to have a strong impact of communication architectures in the smart grid. One example is to develop game theoretic algorithms that can enable quality-of-service guarantees (e.g., in terms of outage or delay) for real-time communications which is crucial in many smart grid applications such as pricing inquiries. In Table~\ref{tab:com}, we show a summary of the applications of game theory for smart grid communications.

Finally, game theory can also be a key enabler for implementing advanced communication architecture such as cognitive radio or cooperative networking within smart grids due to its proven robustness in overcoming classical communication challenges such as interference mitigation, resource allocation, and spectrum sharing~\cite{MC01}.\vspace{-0.4cm}

\begin{table*}[!t] \vspace*{-0.72em}
\caption{Summary of game theoretic techniques for communications in smart grid systems.}\vspace{-0.52cm}
\begin{center}
\begin{tabular}{|p{0.2\textwidth}|p{0.2\textwidth}|p{0.4\textwidth}|}
  \hline
  {\bf Application} &{\bf Game Theoretic Technique}& {\bf Main Future Extensions}\\
  \hline
Subsection~\ref{sec:mhop}: Multi-hop power line communication (such as in \cite{WSPLC01}).& Network formation games &
    \begin{itemize}
  \item  Develop a dynamic model for jointly performing graph formation and frequency allocation.
  \item Study foresighted network formation models for power line communication.
   \item Analyze learning models for addressing practical implementation concerns for network formation games in smart grid systems.
      \end{itemize}
      \\
                    \hline
         \multicolumn{3}{|p{0.8\textwidth}|}{The area of communications in smart grid systems is still at its infancy. Several future opportunities for game theoretic approaches exist such as developing noncooperative techniques for communication protocol selection, study strategic decisions in the presence of security concerns, develop secure routing algorithms using differential games, resource allocation using Nash bargaining, and many other applications.} \\
        \hline
\end{tabular}\label{tab:com}\vspace{-0.8cm}
\end{center}
\end{table*}

\section{Summary}\label{sec:conc}\vspace{-0.1cm}
In this survey, we provided a comprehensive overview on the applications of game theory in smart grid networks. The smart grid applications were carefully drawn from a broad range of problems spanning emerging areas such as micro-grids, demand-side management, and communications. In each area, we have identified the main technical challenges and presented an elaborate discussion on how game theory can be applied to address these challenges. Moreover, we proposed several future directions for extending these approaches and adopting advanced game theoretic techniques, so as to reduce the gap between theoretical models and practical implementations of future smart grids.

Essentially, from the surveyed works, we can clearly note that game theory has a strong potential to provide solutions for pertinent problems in smart grids but also faces many design challenges. However, we also note that many of the existing works have focused on classical static noncooperative games. Hence, for future works, it is of interest to investigate dynamic game models (both in cooperative and noncooperative settings)  and their applications in smart grid systems. The motivation for studying dynamic models stems from the pervasive presence of time-varying parameters in smart grids such as generation, demand, among others. In this context, dynamic game theory could be a cornerstone for capturing these parameters and designing better algorithms for improving the economical and technical aspects of future smart grids.

Beyond dynamic games, it is also of interest to further investigate applications of Bayesian games in smart grids. Bayesian games are a type of noncooperative games in which the players have very limited information on the objective functions and strategies of their opponents. Given the large-scale nature of the smart grid, the involved players in any game model might face several technical difficulties in estimating the exact strategies or objectives of the other players. In this context, future work could investigate how Bayesian games can overcome this difficulty.

Due to space limitation, this article mainly addressed three emerging areas for applying game theory in smart grid networks. However, the use of game theory can easily help overcome several technical challenges in other equally interesting areas in smart grid systems as well. For example, securing cyber-physical systems  that integrate multiple technologies for communications, control, and sensing such as the smart grid is a challenging issue due to the need to introduce security measures at different levels ranging from the communication and infrastructure level to the actual control and power system. To this end, game theory can be used to address the vulnerabilities of smart grids at different levels such as infrastructure, communication and information routing, and state estimation, among others. For example, recent works such as \cite{QZ00} have discussed the use of differential games for securing the smart grid infrastructure given the tradeoff between security and accessibility (in terms of delay and packet drop). Also, the work in \cite{QZ01} sheds a light on how dynamic network formation games can be used to secure the routing of data in the smart grid system. 

Beyond infrastructure and communications, game theory can also be used for securing power grid state estimation against data injection attacks which have received considerable attention recently such as in~\cite{LT00} and references therein. Hence, future work can investigate two important aspects of data injection in smart grids: (i)- the use of dynamic zero-sum noncooperative games for modeling the interactions between a smart grid operator and a data injection attacker, and (ii)- the use of noncooperative games, in conjunction with cooperative games, for studying coordinated data injection attacks  and corresponding defense strategies.

In a nutshell, this paper presented a comprehensive overview on the potential of applying game theory within future smart grid systems. Clearly, game theory will constitute a strong tool for designing future smart grid systems that can fulfill the promise of a completely integrated solution and satisfy the ``sense, communicate, computer, control'' paradigm.\vspace{-0.2cm}

\renewcommand{\baselinestretch}{0.95}
\bibliographystyle{IEEEtran}
\bibliography{references}

\end{document}